\documentclass{SCIS2025}
\usepackage{makecell,multirow,diagbox}
\usepackage{graphicx}

\begin{document}
\ArticleType{RESEARCH PAPER}
\Year{2025}
\Month{}
\Vol{}
\No{}
\DOI{}
\ArtNo{}
\ReceiveDate{}
\ReviseDate{}
\AcceptDate{}
\OnlineDate{}
\AuthorMark{}
\AuthorCitation{}

\title{WiFo: Wireless Foundation Model \\for Channel Prediction}{WiFo: Wireless Foundation Model for Channel Prediction}

\author[1]{Boxun LIU}{}
\author[2]{Shijian GAO}{}
\author[1]{Xuanyu LIU}{}
\author[1]{Xiang CHENG}{xiangcheng@pku.edu.cn}
\author[2,3]{Liuqing Yang}{}


\address[1]{State Key Laboratory of Photonics and Communications, School of Electronics,
Peking University, Beijing 100871, China}
\address[2]{Internet of Things Thrust, The Hong Kong University of
Science and Technology (Guangzhou), Guangzhou 511400, China}
\address[3]{Department of Electronic and Computer Engineering and Department of Civil and Environmental Engineering, \\The Hong Kong University of Science and Technology, Hong Kong, China}

\abstract{Channel prediction permits to acquire channel state information (CSI) without signaling overhead.
However, almost all existing channel prediction methods necessitate the deployment of a dedicated model to accommodate a specific configuration.
{Leveraging the powerful modeling and multi-task learning capabilities of foundation models, we propose the first space-time-frequency (STF) wireless foundation model (WiFo) to address time-frequency channel prediction tasks in a unified manner.}
Specifically, WiFo is initially pre-trained over massive and extensive diverse CSI datasets.
Then, the model will be instantly used for channel prediction under various CSI configurations without any fine-tuning.
We propose a masked autoencoder (MAE)-based network structure for WiFo to handle heterogeneous STF CSI data, and design several mask reconstruction tasks for self-supervised pre-training to capture the inherent 3D variations of CSI.
To fully unleash its predictive power, we build a large-scale heterogeneous simulated CSI dataset consisting of 160K CSI samples for pre-training. 
Simulations validate its superior unified learning performance across multiple datasets and demonstrate its state-of-the-art (SOTA) zero-shot generalization performance via comparisons with other full-shot baselines.
}


\maketitle

\section{Introduction}
Massive multi-input multi-output (MIMO) and orthogonal frequency division multiplexing (OFDM) have been two cornerstone technologies over the past decade \cite{cheng2023intelligent}. 
The performance of MIMO-OFDM systems hinges upon accurate channel state information (CSI), typically obtained through channel estimation \cite{gao2020estimating}.
However, in highly-dynamic scenarios, the channel coherence time is significantly limited, resulting in a sharp increase in channel estimation overhead.
Therefore, channel prediction is proposed as a promising technology to reduce the overhead of CSI acquisition, where the estimated CSI is extrapolated in time or frequency.
Extensive studies have been conducted on model-based channel prediction methods, including traditional autoregressive and polynomial extrapolation methods, as well as advanced vector Prony-based prediction methods \cite{yin2020addressing} and high-resolution parameter estimation schemes \cite{rottenberg2020performance}.
Nevertheless, due to the difficulty of modeling complicated practical channels, the prediction accuracy of model-based approaches is limited.

Deep learning has emerged as a powerful tool to capture complex patterns in a data-driven manner without any prior assumption. 
It has been widely applied in various aspects of wireless communications\cite{wang2022transformer, zhu2023pushing}, including time-domain and frequency-domain channel prediction addressed in this paper.
For instance, several neural networks, including transformer \cite{jiang2022accurate}, ConvLSTM \cite{liu2022spatio}, and variational autoencoders (VAE) \cite{liu2021fire}, have been explored for channel prediction to learn temporal and frequency variations. 
In addition, several channel prediction methods based on physics-informed deep learning have been proposed to incorporate prior knowledge of the CSI or the propagation environment \cite{zhang2024integrated}.
Recently, pre-trained large language models (LLMs) have been applied to MIMO-OFDM channel prediction \cite{liu2024llm4cp} for enhanced accuracy and generalizability.
However, existing deep learning-based channel prediction research is still at the starting stage and awaiting several improvements.
First, due to the limited model scale, existing small model-based methods are not well poised to predict complex practical spatial-temporal-frequency channels accurately.
Secondly, existing deep learning-based schemes are only trained under specific CSI distributions and specific system configurations.
Correspondingly, these methods require retraining when scenarios and system parameters change, which leads to significant training overhead.
Moreover, separate networks are designed and deployed for time-domain and frequency-domain channel prediction, increasing the storage, computation, and management overhead at the base station (BS).

Recently, the emergence of foundational models \cite{bommasani2021opportunities} has led to a paradigm shift in deep learning. 
Specifically, large-scale neural networks pre-trained on vast and diverse data in a self-supervised manner can achieve remarkable generalization capabilities across various downstream tasks, significantly outperforming task-specific models.
Its powerful modeling capabilities and multi-task learning potential offer promise for addressing the limitations of existing channel prediction methods \cite{fontaine2024towards}.
Most recently, a few studies \cite{salihu2024self,alikhani2024large} apply self-supervised pre-training schemes to wireless channel representation learning for CSI-related downstream tasks, such as fingerprint localization and cross-band feature extraction.
Specifically, the pre-trained model is fine-tuned for specific wireless tasks through few-shot learning.
Nevertheless, these cannot be directly adopted for channel prediction.
First, these schemes can only handle space-frequency two-dimensional CSI and cannot address temporal variations.
Secondly, they utilize the encoder-only network instead of an autoencoder, rendering these approaches suitable for classification or regression tasks but less effective for reconstruction tasks such as channel prediction.
Last but not least, they can only reduce, but not eliminate, the retraining overhead for different tasks and scenarios.

Recognizing the limitations of existing studies, we make the pioneering attempt to apply foundation models to channel prediction tasks.
Specifically, we aim to establish a large-scale wireless foundation model pre-trained on extensive CSI data, and directly apply it to various channel prediction tasks under different CSI configurations and scenarios without any fine-tuning.
Evidently, building such a foundational model is quite challenging.
First, it is challenging to address different types of channel prediction tasks with a single network.
Secondly, it is also difficult to use a single network to simultaneously cope with diverse CSI data, where the number of antennas, sub-carriers, and time samples vary.

In order to overcome these challenges, we propose the first wireless foundation model for channel prediction tasks, termed as WiFo.
Different from all existing works, we formulate time-domain and frequency-domain channel prediction tasks as a unified channel reconstruction problem, which captures the complete CSI based on partial CSI.
Inspired by the masked autoencoders (MAE) for image and video self-supervised learning, we propose an MAE-based network structure suitable for CSI reconstruction.
We apply 3D patching and embedding to convert diverse CSI into varying numbers of tokens, enabling efficient processing by the transformer blocks.
For both the encoder and the decoder, we propose a novel positional encoding (STF-PE) structure to learn the 3D position information associated with the CSI.
To capture the inherent 3D variations, we propose three self-supervised pre-training tasks, namely random, time, and frequency-masked reconstruction.
The pre-trained model can be directly applied to zero-shot inference for both channel prediction tasks.
To evaluate the performance of WiFo, we construct a heterogeneous CSI dataset with 16 different configurations of time sampling, sub-carrier, and antenna, containing 160K training samples.
Preliminary results validate its superior multi-dataset unified learning performance and zero-shot generalization capability.
It is also worth emphasizing that its zero-shot prediction performance in unseen scenarios surpasses the full-shot performance of all baselines trained on 10K samples, completely eliminating the retraining or fine-tuning costs. \footnote{Simulation codes are provided to reproduce the results presented in this paper: https://github.com/liuboxun/WiFo}

The primary contributions of our work are summarized as follows.
\begin{itemize}
  \item We propose the first wireless foundation model (WiFo) designed to uniformly facilitate time-domain and frequency-domain channel prediction tasks. 
  {To the best of our knowledge, it is the first versatile model capable of simultaneously tackling different types of channel prediction tasks under diverse CSI configurations.}
  \item We develop an MAE-based network structure to cope with heterogeneous CSI data and introduce three mask reconstruction tasks for self-supervised pre-training, aimed at capturing the inherent space-time-frequency correlations of CSI. 
  The pre-trained model can be directly utilized for inference without the need for fine-tuning.
  \item We construct 16 diverse datasets with various CSI configurations using the QuaDRiGa channel generator for pre-training, consisting of 160K training samples. Simulations confirm that WiFo can effectively learn across different channel prediction tasks and heterogeneous datasets and demonstrates strong zero-shot performance in new scenarios.
\end{itemize}

\textit{Notation}:  $\Vert\cdot\Vert_F$ denotes the Frobenius norm. 
$\bm{a}[i]$ is the $i\mbox{-}$th element of a vector $\bm{a}$ and $\bm{A}[i,j]$ denotes the element of matrix $\bm{A}$ at the $i\mbox{-}$th row and the $j\mbox{-}$th column. 
$\mathbb{R}$ and $\mathbb{C}$ denote the set of real numbers and complex numbers, respectively.

\section{System Model and Problem Formulation}
In this paper, we consider a multi-input single-output (MISO)-OFDM system, where the BS and the user are equipped with a uniform planar array (UPA) and a single antenna, respectively.
The UPA contains $N=N_h \times N_v$ elements, with $N_h$ and $N_v$ being the number of antennas along the horizontal and vertical directions, respectively.
{The proposed scheme can be directly extended to the case of multi-user channel prediction by processing multiple samples in parallel.}

\subsection{3D Channel Model}\label{3D channel model}
We adopt the classical geometric channel model \cite{zhang2024latest} consisting of $P$ paths.
The CSI between the BS and the user sampled at frequency $f$ and time $t$ can be expressed as
\begin{align}
    \bm{h}(t,f)=\sum_{p=1}^{P}g_p\bm{a}(\phi_p,\theta_p)e^{-j2\pi f\tau_p}e^{j2\pi \nu_p t},
\end{align}
where $g_p$, $\tau_p$, and $\nu_p$ represent the complex amplitude, the delay, and the Doppler shift associated with the $p$-th path.
$\bm{a}(\phi_p,\theta_p)\in \mathbb{C}^{N \times 1}$ is the steering vector of UPA\cite{liu2024llm4cp}, where $\phi_p$ and $\theta_p$ denote the azimuth and elevation angles, respectively.

Assume that the considered time-frequency region \cite{liu2024llm4cp} spans $T$ and $K$ resource blocks (RBs) along the time and frequency dimensions, respectively.
The pilot placement pattern is identical for all antennas and each RB contains a pilot.
The pilot intervals along time and frequency are $\Delta t$ and $\Delta f$, respectively.
Without loss of generality, we consider the CSI only at the pilot positions.
The considered space-time-frequency CSI is denoted as $\bm{H}\in \mathbb{C}^{T \times K \times N}$, which satisfies
\begin{align}
    \bm{H}[i,j,:]=\bm{h}(i\Delta t,f_1+(j-1)\Delta f),\quad i = 1, \dots, T, \quad j = 1, \dots, K,
\end{align}
where $f_1$ represents the frequency at the pilot position of the first RB in the frequency domain.

\subsection{Problem Description}\label{JPF}
We consider two types of channel prediction tasks: time-domain prediction and frequency-domain prediction. 

1) \textbf{Time-domain channel prediction}:
We aim to predict CSI for the future $T-T_h$ RBs based on the historical $T_h$ RBs along the time dimension.
Denote the mapping function of time-domain channel prediction as $\Phi_{\rm t}$, the prediction process is derived as
\begin{align}\label{time-pre}
    \bm{H}[T_h+1:T,:,:] = \Phi_{\rm t}(\bm{H}[1:T_h,:,:]).
\end{align}

2) \textbf{Frequency-domain channel prediction}: 
We focus on channel prediction for adjacent frequency bands.
Without loss of generality, we aim to predict the last $K-K_u$ RBs via the first $K_u$ RBs along the frequency dimension. 
Such an operation can be represented as
\begin{align}
    \bm{H}[:,K_u+1:K,:] = \Phi_{\rm f}(\bm{H}[:,1:K_u,:]),
\end{align}
where $\Phi_{\rm f}$ is the mapping function in the frequency-domain.

Previous studies have designed separate networks to manage individual prediction tasks and various CSI configurations. However, since a BS must simultaneously handle multiple channel prediction tasks and diverse user configurations, this approach leads to significant overhead in network deployment. To address this issue, we propose to unify the channel prediction tasks into a single channel reconstruction task using our wireless foundation model. Specifically, the formulated channel reconstruction task seeks to derive the complete CSI from partial CSI. We denote the partial CSI as $H[\Omega]$, where $\Omega$ represents the subset of all elements. Our goal is to develop a reconstruction function  $\Phi_{\rm rec}$ that facilitates a universal mapping as follows:
\begin{align}\label{mapping}
    \bm{H} = \Phi_{\rm rec}({\bm{H}}[\Omega]),
\end{align}
where $\Omega$ represents the temporal or frequency subset for time-domain and frequency-domain channel prediction, respectively.
It is worth noting that the function $\Phi_{\rm rec}$ needs to handle diverse CSI data with arbitrary $T$, $K$, and $N$, while simultaneously performing reconstruction in the time or frequency domain.
However, traditional deep learning-based channel prediction schemes and LLM-empowered schemes \cite{liu2024llm4cp} are designed for fixed-size CSI and specific prediction tasks, which limits their ability to achieve this universal mapping.
{For instance, most existing deep-learning-based prediction schemes design neural networks under fixed input dimensions, where the adopted fully connected layers are incapable of handling a varying input size.}
Consequently, we aim to establish a foundation model to address this challenge.

\begin{figure*}[t]
\center{\includegraphics[width=16cm]  {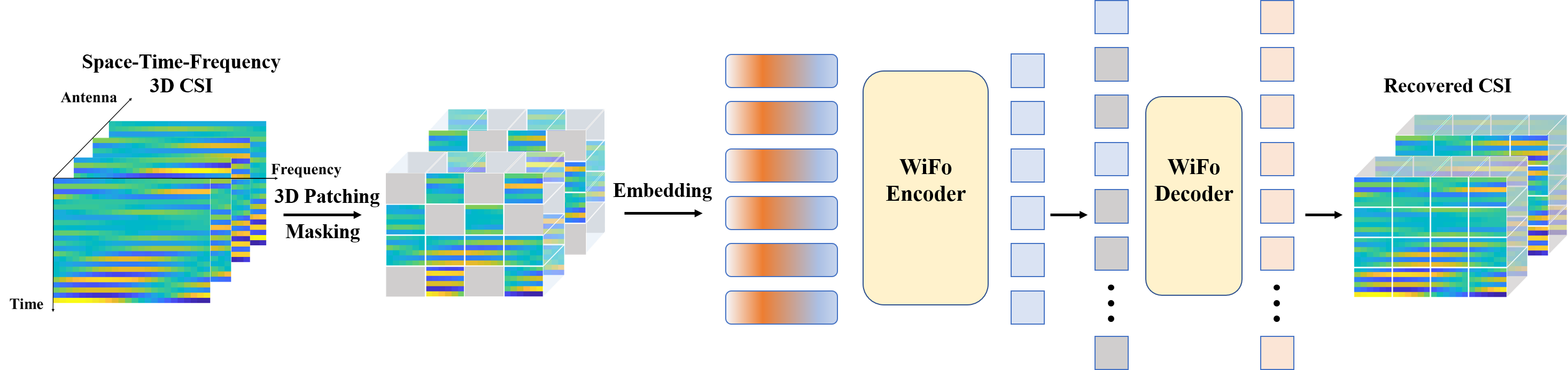}} 
\caption{{An illustration of the proposed WiFo: the original space-time-frequency 3D CSI is firstly divided into 3D patches and masked, with the visible patches embedded as CSI tokens. The WiFo encoder processes the input tokens, and after concatenating the learnable mask tokens, the WiFo decoder reconstructs the original 3D CSI.}}
 \label{Network}
\end{figure*}
\section{Wireless Foundation Model}
In this section, a novel wireless foundation model, termed WiFo, is proposed to realize the universal channel prediction.
First, the network structure of WiFo is introduced based on masked autoencoders (MAE).
Then, a self-supervised pre-training scheme is proposed based on multiple reconstruction tasks.
Finally, the pre-trained WiFo is directly applied for channel prediction without any fine-tuning.

\subsection{Network Structure}
MAE \cite{he2022masked,tong2022videomae,feichtenhofer2022masked} has been demonstrated to be a data-efficient self-supervised pre-training framework for image and video representation learning.
It adopts an asymmetric encoder-decoder architecture, where an encoder operates on unmasked tokens and a lightweight decoder recovers the original image or video.
Inspired by MAE, WiFo adopts a transformer-based encoder-decoder network structure, as shown in Fig. \ref{Network}.
It consists of four major blocks, namely CSI embedding, masking, encoder, and decoder. Each of them will be described in detail below.

\subsubsection{CSI Embedding}
To facilitate neural network processing, the complex $\bm{H}$ is first converted into a real-valued tensor $\tilde{\bm{H}}\in \mathbb{R}^{2\times T \times K \times N}$, which consists of two channels for the real and imaginary parts.
To convert $\tilde{\bm{H}}$ into 1D sequential data suitable for transformer processing, we apply 3D patching \cite{tong2022videomae} to the last three dimensions.
Specifically, the size of each 3D patch is $(t,k,n)$, where $t$, $k$, and $n$ represent the patch size along the time, frequency, and space dimensions.
Then the non-overlapping 3D patches, with a total number of $ L=\frac{T}{t} \times \frac{K}{k} \times \frac{N}{n}$, are flattened and embedded into a series of tokens with dimension $D_{\rm enc}$, where $D_{\rm enc}$ is the hidden size of the encoder.
The above operation can be implemented using 3D convolution, i.e.,
\begin{align}
\bm{H}_{\rm conv}=Conv3d(\tilde{\bm{H}}),
\end{align}
where $Conv3d(\cdot)$ represents the 3D convolution operator with 2 input channels and $D_{\rm enc}$ output channels.
Then the convolution result $\bm{H}_{\rm conv}\in \mathbb{R}^{D_{\rm enc} \times \frac{T}{t}\times \frac{K}{k} \times \frac{N}{n}}$ is flattened into $\bm{H}_{\rm emb}\in \mathbb{R}^{D_{\rm enc} \times L}$ as the CSI tokens.

\subsubsection{Masking} 
As shown in Fig. \ref{Network}, several 3D patches are masked. 
It is equivalent to mask partial tokens of $\bm{H}_{\rm emb}$ and retains only a subset of tokens to be processed by the encoder module.
Denote the visible tokens and the masked tokens as $\bm{H}_{\rm vis}\in \mathbb{R}^{D_{\rm enc} \times L_{\rm vis}}$ and $\bm{H}_{\rm mask}\in \mathbb{R}^{D_{\rm enc} \times (L-L_{\rm vis})}$, respectively, where $L_{\rm vis}$ is the number of visible tokens.
Then the masking process is represented as
\begin{align}
[\bm{H}_{\rm vis},\bm{H}_{\rm mask}]=Mask(\bm{H}_{\rm emb}),
\end{align}
where $Mask(\cdot)$ represents the masking operation with a certain masking strategy and masking ratio.

The role of the masking operation differs between the model pre-training and inference stages.
During the pre-training stage, masking strategies are performed as self-supervised pre-training reconstruction tasks for better CSI representation learning.
During the inference stage, masking strategies are performed to apply the proposed model to specific channel prediction tasks.

\begin{figure}[t]
\center{\includegraphics[width=14cm]  {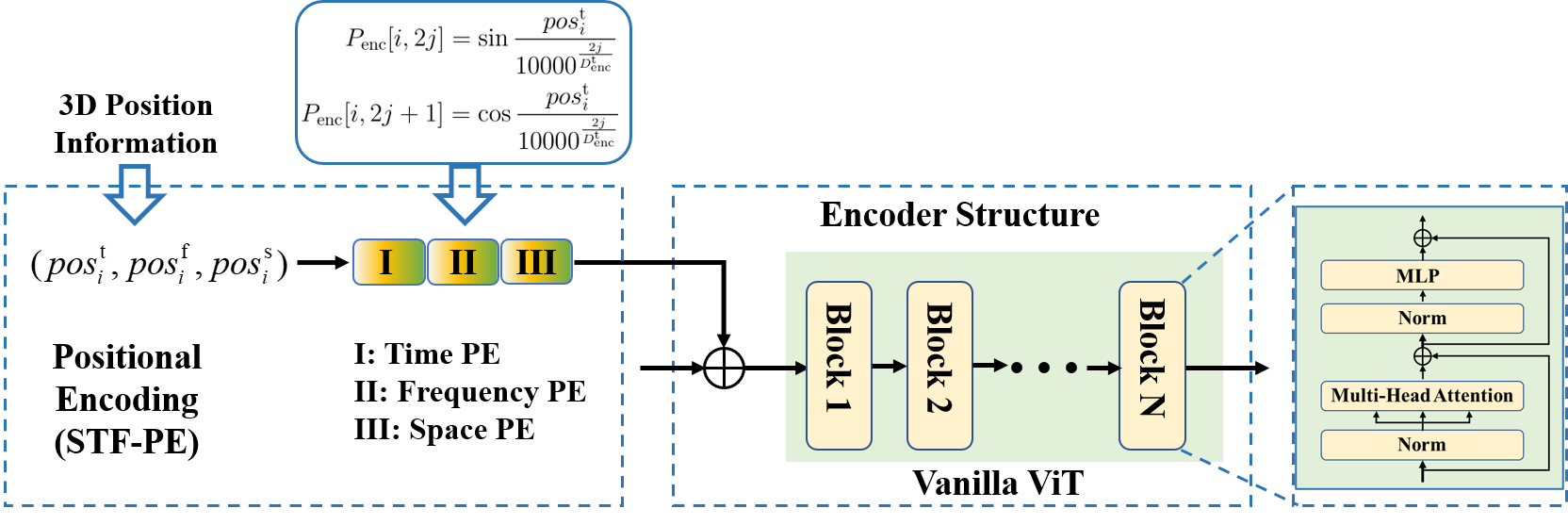}} 
\caption{An illustration of the architecture of the encoder module. }
 \label{Encoder}
\end{figure}
\subsubsection{Encoder}
The architecture of the encoder is shown in Fig. \ref{Encoder}.
Consistent with MAE \cite{he2022masked,tong2022videomae,feichtenhofer2022masked}, the visible tokens are sequentially added with positional encoding and processed by a series of transformer blocks implemented by vanilla Vision Transformer (ViT) backbone \cite{dosovitskiy2020image}.
Denote the operation of the transformer blocks as $f_{\rm enc}$, the encoder output $\bm{H}_{\rm enc}\in \mathbb{R}^{D_{\rm enc} \times L_{\rm vis}}$ is derived as 
\begin{align}
\bm{H}_{\rm enc}=f_{\rm enc}(\bm{H}_{\rm vis}+\bm{P}_{\rm enc}),
\end{align}
where $\bm{P}_{\rm enc}\in \mathbb{R}^{D_{\rm enc} \times L_{\rm vis}}$ denotes the positional encoding.

To enable the model to learn the three-dimensional position information of CSI, we propose an STF positional encoding (STF-PE). 
Specifically, separate positional encoding for time, frequency, and space, denoted as $\bm{P}_{\rm enc}^{\rm t}\in \mathbb{R}^{ L_{\rm vis}\times D_{\rm enc}^{\rm t}}$, $\bm{P}_{\rm enc}^{\rm f}\in \mathbb{R}^{ L_{\rm vis}\times D_{\rm enc}^{\rm f}}$, and $\bm{P}_{\rm enc}^{\rm s}\in \mathbb{R}^{ L_{\rm vis}\times D_{\rm enc}^{\rm s}}$, are concatenated along the feature dimension, satisfying $D_{\rm enc}= D_{\rm enc}^{\rm t} + D_{\rm enc}^{\rm f} + D_{\rm enc}^{\rm s}$.
Without loss of generality, let $D_{\rm enc}^{\rm t}=D_{\rm enc}^{\rm f}=\lfloor D_{\rm enc}/3 \rfloor$ and $D_{\rm enc}^{\rm s}=D_{\rm enc} - 2\lfloor D_{\rm enc}/3 \rfloor$, respectively.
For better generalization across different sizes of CSI, each separate positional encoding of STF-PE adopts absolute SinCos positional encoding \cite{he2022masked} instead of learnable encoding\cite{feichtenhofer2022masked}.
As an example, consider the time positional encoding. 
For $i$-th visible token, denote the temporal, spatial, and frequency coordinates of its corresponding 3D patch as $(pos_i^{\rm t},pos_i^{\rm f},pos_i^{\rm s})$, we have:
\begin{align}
\bm{P}_{\rm enc}^{\rm t}[i,2j]=\sin{\frac{pos_i^{\rm t}}{10000^{\frac{2j}{D_{\rm enc}^{\rm t}}}}},\ {\rm and} \ \bm{P}_{\rm enc}^{\rm t}[i,2j+1]=\cos{\frac{pos_i^{\rm t}}{10000^{\frac{2j}{D_{\rm enc}^{\rm t}}}}}.
\end{align}

\subsubsection{Decoder}
The decoder is designed to reconstruct the original $\bm{H}$ from the encoder output.
$\bm{H}_{\rm enc}$ is first converted to $\bar{\bm{H}}_{\rm enc}\in \mathbb{R}^{D_{\rm dec} \times L_{\rm vis}}$ via a fully connected layer to align with the feature dimension $D_{\rm dec}$ of the decoder transformer blocks.
$\bar{\bm{H}}_{\rm enc}$ is then concatenated with learnable mask tokens $\bm{M}\in \mathbb{R}^{D_{\rm dec} \times (L-L_{\rm vis})}$, and added with decoder positional encoding $\bm{P}_{\rm dec}\in \mathbb{R}^{D_{\rm dec} \times L}$ before processed by lightweight ViT transformer blocks, denoted as $f_{\rm dec}$.
$\bm{P}_{\rm dec}$ adopts the same STF-PE as the encoder.
Denote the output of the decoder transformer blocks as $\bm{H}_{\rm dec}\in \mathbb{R}^{D_{\rm dec} \times L}$, and we have
\begin{align}
\bm{H}_{\rm dec}=f_{\rm dec}([\bm{H}_{\rm vis},\bm{M}]+\bm{P}_{\rm dec}).
\end{align}
Then, $\bm{H}_{\rm dec}$ is transformed into $\tilde{\bm{H}}_{\rm pred}\in \mathbb{R}^{2\times T \times K \times N}$ by a fully connected layer and reshape operation.
Finally, we obtain the reconstructed complex CSI $\bm{H}_{\rm pred}\in \mathbb{C}^{T \times K \times N}$, i.e.,
\begin{align}
\bm{H}_{\rm pred}=\tilde{\bm{H}}_{\rm pred}[1,:,:,:]+1j\times \tilde{\bm{H}}_{\rm pred}[2,:,:,:].
\end{align}

\subsection{Self-Supervised Pre-Training}
\begin{figure}[t]
\center{\includegraphics[width=13cm]  {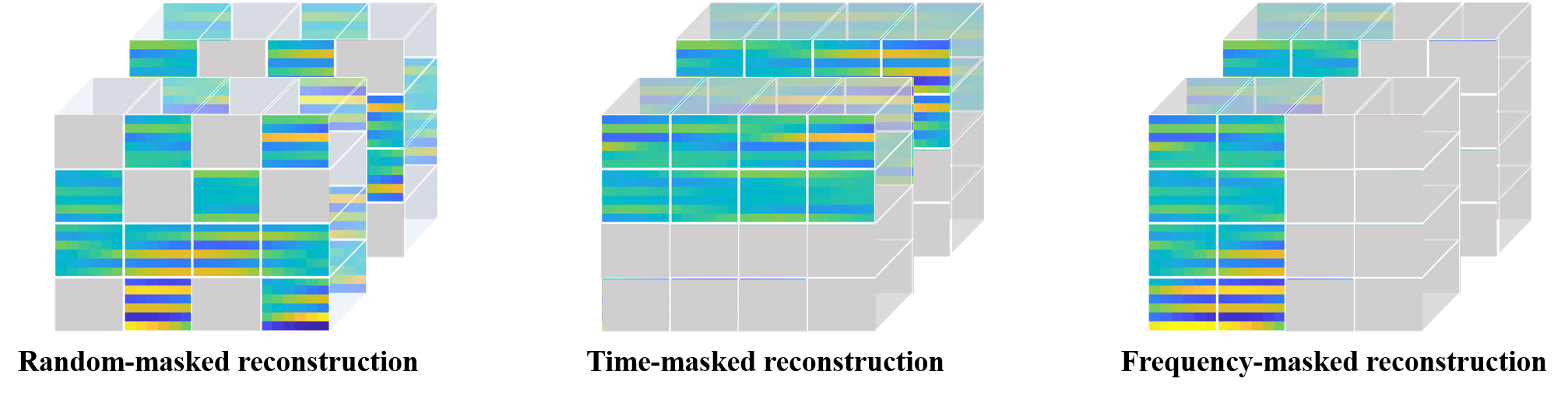}} 
\caption{An illustration of the proposed masked reconstruction tasks. }
 \label{mask_fig}
\end{figure}
Unlike existing channel and prediction schemes \cite{jiang2022accurate, liu2022spatio,liu2021fire,liu2024llm4cp} which are trained and tested on a specific dataset, the proposed WiFo is first pre-trained across multiple heterogeneous datasets and then directly applied to unseen scenarios and new CSI configurations.
Specifically, several self-supervised training tasks are designed to capture the intricate inherent space-time-frequency correlations of CSI.
Masked reconstruction \cite{he2022masked,tong2022videomae,feichtenhofer2022masked}  has been proven to be an effective pre-training task for downstream visual tasks, which masks random patches of an image and then reconstructs the complete image.
Notably, temporal-domain and frequency-domain channel prediction are special types of masked reconstruction tasks, where patches are masked along the time or frequency dimensions.
Therefore, we propose three masked reconstruction tasks to capture the intricate inherent space-time-frequency correlations of CSI, as shown in Fig. \ref{mask_fig}.
Their detailed operations are introduced as follows.
To simplify the description, we denote the temporal, spatial, and frequency coordinates of the 3D patch corresponding to the $i$-th token in $\bm{H}_{\rm emb}$ as $(\overline{pos}_{i}^{\rm t}, \overline{pos}_{i}^{\rm f},\overline{pos}_{i}^{\rm s})$. 

1) \textbf{Random-masked reconstruction}: 
It randomly masks tokens from all tokens with a $R_{\rm r}$ ratio. 
This masking strategy is isotropic across the space-time-frequency dimensions to capture the 3D structured features.

2) \textbf{Time-masked reconstruction}: 
It is designed to enhance the time-domain channel prediction task.
In this masking strategy, future tokens are masked with a certain ratio $R_{\rm t}$, i.e., all tokens with $\overline{pos}_{i}^{\rm t}\ge(\frac{T}{t}-\lfloor R_{\rm t}\frac{T}{t}\rfloor)$ are masked.
It helps the model learn the causal relationships of CSI over time.

3) \textbf{Frequency-masked reconstruction}:
This task is designed to improve Wifo's performance in frequency-domain channel prediction.
Unlike the time-masked reconstruction, frequency masking strategy masks tokens along frequency dimension with a ratio $R_{\rm f}$, so all tokens with $\overline{pos}_{i}^{\rm f}\ge(\frac{K}{k}-\lfloor R_{\rm f}\frac{K}{k}\rfloor)$ are masked.
This helps the model learn the variations between adjacent frequency bands.

During self-supervised pre-training, the three pre-training tasks mentioned above are executed sequentially for each batch.
Our objective is to minimize the reconstruction error, for which we adopt the mean squared error (MSE) as the loss function.
The loss is computed using only the reconstructed CSI points and is defined as
\begin{align}
\mathcal{L}=\frac{1}{\left|\omega\right|}\Vert\bm{H}[\omega]-\bm{H}_{\rm pred}[\omega]\Vert_F^2,
\end{align}
where $\omega$ represents the temporal or frequency predicted CSI subset.

\subsection{Model Inference}
Once the pre-training stage is complete, the pre-trained WiFo can perform zero-shot inference on both the time-domain and frequency-domain tasks, as shown in Eq. \ref{mapping}.
Specifically, the portion to be predicted is first padded with zeros, and then the zero-padded CSI is fed into WiFo.
Subsequently, the corresponding time-masked or frequency-masked strategy is applied with a certain masking ratio, allowing the network to output the reconstructed part as the predicted CSI.
{Below, we will elaborate on the process in detail, using time-domain channel prediction as an example.}

{As shown in Eq. \ref{time-pre}, we aim to predict future CSI $\bm{H}[T_h+1:T,:,:]$ based on historical CSI $\bm{H}[1:T_h,:,:]$.
First, the historical CSI is zero-padded into $\bm{H}_{\rm in}\in \mathbb{C}^{T \times K \times N}$ to be fed into WiFo, satisfying
\begin{align}
\bm{H}_{\rm in}[1:T_h,:,:]=\bm{H}[1:T_h,:,:]\ \  {\rm and}\  \bm{H}_{\rm in}[T_h+1:T,:,:]=\bm{0}_{(T-T_h)\times K\times N},
\end{align}
where $\bm{0}_{(T-T_h)\times K\times N}$ represents a all-zero tensor of dimensional $(T-T_h)\times K\times N$.
Then the time-domain masking is applied, i.e., all tokens with $\overline{pos}_{i}^{\rm t}\ge T_h+1$ are masked.
By denoting the reconstructed CSI output of WiFo as $\bm{H}_{\rm out}$, the predicted future CSI can be obtained as $\bm{H}_{\rm pre}=H_{\rm out}[T_h+1:T,:,:]$.}

\section{Experiments}
In this section, we conduct extensive experiments to demonstrate the effectiveness of the proposed WiFo.
First, we built a series of simulated datasets covering various system configurations for network pre-training and inference.
Then, experimental setups are introduced, including network and training parameters, baselines, and the performance metric.
Finally, the channel prediction performance of the proposed WiFo is comprehensively evaluated.

\subsection{Datasets}
\begin{table}[t]
\footnotesize
\caption{An illustration of the system configurations of the constructed 3D CSI datasets.}\label{Dataset}
\centering
\begin{tabular}{ccccccccc}
\toprule[1pt]
Dataset & $f_{\rm C}$(GHz) & $K$ & $\Delta f$(kHz) & T & $\Delta t$(ms) & UPA & Scenario & User speed(km/h) \\ \midrule[1pt]
    D1  &  1.5   &  128 & 90  &    24     & 1  &    $1\times 4$     &    UMi+NLoS &  3-50    \\ \hline
    D2  &  1.5   &  128 & 180  &    24     & 0.5  &    $2\times 4$     &    RMa+NLoS  &  120-300   \\ \hline
    D3  &  1.5   &  64 & 90  &    16     & 1  &    $1\times 8$     &    Indoor+LoS  & 0-10    \\ \hline
    D4  &  1.5   &  32 & 180  &    16     & 0.5  &    $4\times 8$     &    UMa+LoS  & 30-100    \\ \hline
    D5  &  2.5   &  64 & 180  &    24     & 0.5  &    $2\times 2$     &    RMa+NLoS  &  120-300   \\ \hline
    D6  &  2.5   &  128 & 90  &    24     & 1  &    $2\times 4$     &    UMi+LoS & 3-50     \\ \hline
    D7  &  2.5   &  32 & 360  &    16     & 0.5  &    $4\times 8$     &    UMa+LoS  &  30-100   \\ \hline
    D8  &  2.5   &  64 & 90   &    16     & 1  &    $4\times 4$     &    Indoor+NLoS  &  0-10    \\ \hline
    D9  &  4.9   &  128 & 180   &    24     & 1  &    $1\times 4$     &    UMi+NLoS &  3-50    \\ \hline
    D10  &  4.9   &  64 & 180   &    24     & 0.5  &    $2\times 4$     &    RMa+LoS  &  120-300   \\ \hline
    D11 &  4.9   &  64 & 90   &    16     & 0.5  &    $4\times 4$     &    UMa+NLoS  &  30-100   \\ \hline
    D12  &  4.9   &  32 & 180   &    16     & 1  &    $4\times 8$     &    Indoor+LoS  & 0-10    \\ \hline
    D13  &  5.9   &  64 & 90   &    24     & 0.5  &    $2\times 8$     &    RMa+LoS  & 120-300    \\ \hline
    D14  &  5.9   &  128 & 180   &    24     & 1  &    $2\times 4$     &    UMi+NLoS & 3-50     \\ \hline
    D15  &  5.9   &  64 & 90   &    16    & 1  &    $4\times 4$     &    Indoor+LoS  & 0-10    \\ \hline
    D16  &  5.9   &  32 & 360   &    16     & 0.5  &    $4\times 8$     &    UMa+NLoS  &  30-100   \\ \midrule[1pt]
    D17  &  3.5   &  32 & 180   &    16     & 0.5  &    $4\times 8$     &    UMa+NLoS & 30-100     \\ \hline
    D18  &  6.7   &  64 & 180   &    24    & 1  &    $4\times 4$     &    UMi+LoS  & 3-50    \\ \hline
{D19} &{28}&{32}&{360}&    {16}    & {0.25}&    {$4\times 8$}     &    {UMa+LoS}  & {30-100}    \\ \bottomrule[1pt]
\end{tabular}
\end{table}
To fully unleash WiFo's prediction capabilities across various CSI configurations, we have constructed a series of diverse 3D CSI datasets, generated through channel generator QuaDRiGa \cite{jaeckel2014quadriga} compliant with the 3GPP standards.
Consistent with our system model in Section \ref{3D channel model}, we consider MISO-OFDM systems, where the BS is equipped with a UPA and the user has a single antenna.
The adjacent antenna spacing is half the wavelength at the central frequency.
{A total of 19 datasets are simulated, indexed from D1 to D19, covering various space-time-frequency CSI configurations, cell scenarios, and user speed.}
Among them, the first 16 datasets are used for pre-training, and the last three datasets are used for generalization testing.
The detailed simulation configurations of each dataset are shown in Table \ref{Dataset}, where $f_{\rm C}$ represents the center frequency.
There are seven 5G New Radio (NR) frequency bands, eight 3GPP \cite{3gpp2018study} scenarios, and seven user speed ranges considered.
For each CSI sample, the user has a random initial position and a straight-line motion trajectory, where the speed is uniformly selected within the corresponding speed range.
Each dataset contains 12000 samples, which are randomly split into 9000, 1000, and 2000 samples for training, validation, and inference, respectively.
All CSI samples are pre-standardized using the mean and variance of the corresponding dataset.
To simulate the imperfect factors of practical CSI acquisition, complex Gaussian noise with 20 dB is added to the CSI samples during both the training and inference process.

\subsection{Experimental Settings} 
We consider both the time-domain and the frequency-domain channel prediction tasks. 
For time-domain channel prediction, we predict the CSI of the future $\frac{T}{2}$ RBs according to historical $\frac{T}{2}$ RBs along the time dimension.
For frequency-domain channel prediction, we predict the CSI of the last $\frac{K}{2}$ RBs using the first $\frac{K}{2}$ RBs as input along the frequency dimension.
\subsubsection{Network and Pre-training Settings}
To investigate the impact of model size on performance, we consider WiFo of 5 different sizes, with the specific parameters listed in Table \ref{Network para}.
In the experiments, we set the size of each 3D patch as $(4,4,4)$.
In addition, we set the masking ratio of random masking, time-domain masking, and frequency-domain masking as $R_{\rm r}=85\%$, $R_{\rm t}=50\%$, and $R_{\rm f}=50\%$, respectively.

We conduct all experiments on the same machine with 4 NVIDIA GeForce RTX4090 GPUs, AMD EPYC 7763 64-Core CPU, and 256 GB of RAM. WiFo and other baselines are trained with TF32 (TensorFloat 32) precision. The pre-training settings are illustrated in Table \ref{Training para}. 
During the pre-training process, the 16 datasets are split into batches with a certain batch size and shuffled.
For each batch, the proposed three masked reconstruction tasks are applied sequentially.
The final loss value for gradient descent is the mean loss of the three reconstruction tasks.

\begin{table}[t]
\footnotesize
\caption{Network parameters of WiFo with different sizes.}\label{Network para}
\centering
\begin{tabular}{cccccccc}
\toprule[1pt]
Model & Enc. depth  & Enc. width & Enc. heads & Dec. depth & Dec. width & Dec. heads & Parameters  \\ \midrule[1pt]
WiFo-Tiny  &  6   &  64 & 8  &  4  & 64 & 8 &  0.3M \\ \hline
WiFo-Little  &  6   & 128 & 8  &  4  & 128 & 8 &  1.4M \\ \hline
WiFo-Small &  6   & 256 & 8  &  4  & 256 & 8 & 5.5M \\ \hline
WiFo-Base  &  6   & 512 & 8  &  4  & 512 & 8 &  21.6M   \\ \hline
WiFo-Large  &  8  & 768 & 8  &  4  & 768 & 8 &  86.1M \\ \bottomrule[1pt]
\end{tabular}
\end{table}

\begin{table}[t]
\footnotesize
\caption{Pre-training parameters of WiFo.}\label{Training para}
\centering
\begin{tabular}{cc}
\toprule[1pt]
Parameter & Value  \\ \midrule[1pt]
Optimizer  &  AdamW ($\beta_1=0.9$,$\beta_1=0.999$, weight decay=0.05)  \\ \hline
Batch size  &  128   \\ \hline
Epochs &  200  \\ \hline
Learning rate schedule   &  Cosine decay \cite{loshchilov2016sgdr} (warmup epochs = 5)   \\ \hline
Base learning rate  &  $5\times10^{-4}$  \\ \bottomrule[1pt]
\end{tabular}
\end{table}

\subsubsection{Baselines}

For a comprehensive comparison, we provide the following five baselines, covering model-based, traditional deep learning-based, and advanced LLM-powered methods.

1) \textbf{PAD}\cite{yin2020addressing}: PAD is a Prony-based angular-delay domain channel prediction method. It is only applied to time-domain prediction tasks. In our experiments, the predictor order is set as $N=4$ and $6$ for the case of $T=16$ and $24$, respectively.

2) \textbf{Transformer}\cite{jiang2022accurate}: The transformer-based prediction scheme is proposed in \cite{jiang2022accurate} for parallel channel prediction. For a fair comparison, it adopts the same CSI embedding method as WiFo. Specifically, 3D patching and embedding are first applied to the original CSI, and the embedded tokens are processed by the network. The feature dimension is set as 128, while the depth of the encoder and decoder are set as 5 and 8, respectively. 

3) \textbf{LSTM}\cite{jiang2020deep}: 
Long short-term memory (LSTM) is proposed for sequential processing to overcome the vanishing gradient problem. In our experiments, we consider a two-layer LSTM. For time-domain channel prediction, antenna and frequency dimensions are flattened and input into the network. For frequency-domain channel prediction, time and antenna dimensions are flattened similarly.

4) \textbf{3D ResNet} \cite{feichtenhofer2019slowfast}: As a network specifically designed for handling 3D data, a ResNet-style model \cite{feichtenhofer2019slowfast} proposed for video recognition tasks is considered for comparison. It consists of 50 weighted layers to capture the three-dimensional relationships within the CSI. 

5) \textbf{LLM4CP}\cite{liu2024llm4cp}:
LLM4CP is a LLM-empowered channel prediction scheme, where GPT-2 is fine-tuned for cross-modality knowledge transfer.
Since the original LLM4CP method cannot directly handle 3D CSI, we consider two implementation approaches.
To ensure a fair comparison, the first approach considers the same 3D patching method as WiFo, termed LLM4CP.
In addition, the second implementation approach adopts antenna parallel processing \cite{liu2024llm4cp}, termed LLM4CP*.

\subsubsection{Performance Metric}
In our experiments, normalized mean square error (NMSE) is adopted to measure the prediction accuracy directly.
Let $\bm{H}_{\rm P}$ and $\bm{H}_{\rm GT}$ denote the predicted part of the CSI and its corresponding ground truth, respectively.
The performance metric NMSE is derived as
\begin{align}
{\rm NMSE}(\bm{H}_{\rm P},\bm{H}_{\rm GT})=\frac{\Vert \bm{H}_{\rm P}-\bm{H}_{\rm GT} \Vert_F^2}{\Vert \bm{H}_{\rm GT} \Vert_F^2}.
\end{align}

\subsection{Performance Evaluation}
\subsubsection{Multi-Dataset Unified Learning}
\begin{table}[t]
\footnotesize
\caption{The NMSE performance of WiFo-Base and other baselines on the time-domain channel prediction task across the D1-D16 datasets. The best results are highlighted in \textbf{bold}, while the second-best results are \underline{underlined}.}\label{D1-D16-T}
\centering
\begin{tabular}{cccccccc}
\toprule[1pt]
Dataset & WiFo-Base & Transformer & LSTM & 3D ResNet & PAD & LLM4CP & LLM4CP*  \\ \midrule[1pt]
    D1  &  \underline{0.082}   &  0.112 & 0.356  &    0.088  & 0.529 & 0.117 & \textbf{0.074}  \\ \hline
    D2  &  \textbf{0.260}   &  0.416 & 0.797  &    0.351  & 1.074 & 0.451 & \underline{0.305}   \\ \hline
    D3  &  0.016   &  0.016 & 0.027  &    \underline{0.014}  & 0.038 & 0.015 & \textbf{0.013}   \\ \hline
    D4  &  \textbf{0.048}   &  0.107 & 0.418  &    0.055  & 0.317 & 0.106 & \underline{0.060}  \\ \hline
    D5  &  \textbf{0.494}   &  0.638 & 0.788  &    0.751  & 5.008 & 0.637 & \underline{0.510} \\ \hline
    D6  &  \textbf{0.095}   &  0.174 & 0.542  &    0.157  & 0.568 & 0.206 & \underline{0.133}  \\ \hline
    D7  &  \textbf{0.081}   &  0.219 & 0.576  &    \underline{0.103}  & 0.617 & 0.198 & 0.112   \\ \hline
    D8  &  \underline{0.018}   &  0.024 & 0.092  &    \textbf{0.016}  & 0.073 & 0.025 & \textbf{0.016}  \\ \hline
    D9  &  0.347   &  0.483 & 0.835  & \underline{0.349}  & 1.087 & 0.475 & \textbf{0.312}  \\ \hline
    D10  &  \textbf{0.467}  & 0.649 & 0.689  & 0.869  & 3.863 & 0.709 & \underline{0.563}  \\ \hline
    D11 &  \textbf{0.227}   &  0.440 & 0.834  &    \underline{0.274}  & 1.017 & 0.405 & \underline{0.273}   \\ \hline
    D12  &  \textbf{0.023}  & 0.035 & 0.166  &    \underline{0.025}  & 0.132 & 0.035 & 0.026 \\ \hline
    D13  &  \textbf{0.482}   & 0.718 & 0.876  &   0.815 & 5.213 & 0.758 & \underline{0.648}  \\ \hline
    D14  &  \underline{0.369}   & 0.546 & 0.884  & 0.388  & 1.021  & 0.562 &  \textbf{0.358}   \\ \hline
    D15  &  \textbf{0.029}   & 0.039 & 0.156  &    0.032  & 0.151  & 0.038 & \underline{0.030}   \\ \hline
    D16  &  \textbf{0.318}   & 0.591 & 0.944  &  0.329   & 1.034  & 0.545 & \underline{0.349}  \\ \midrule[1pt]
    Average  & \textbf{0.210}   & 0.325  & 0.561 &  0.289    & 1.359  & 0.330 & \underline{0.236}   \\ \bottomrule[1pt]
\end{tabular}
\end{table}
\begin{table}[th]
\footnotesize
\caption{The NMSE performance of WiFo-Base and other baselines on the frequency-domain channel prediction task across the D1-D16 datasets. The best results are highlighted in \textbf{bold}, while the second-best results are \underline{underlined}.}\label{D1-D16-F}
\centering
\begin{tabular}{ccccccc}
\toprule[1pt]
Dataset & WiFo-Base & Transformer & LSTM & 3D ResNet & LLM4CP & LLM4CP*  \\ \midrule[1pt]
    D1  &  \textbf{0.318}   &  0.532 & 0.705  &    0.839  & 0.392 & \underline{0.375}       \\ \hline
    D2  &  \textbf{0.181}   &  0.556 & 0.763  &    0.647  & 0.419 & \underline{0.223}       \\ \hline
    D3  &  0.027   &  \textbf{0.016} & 0.037  &    0.071  & \underline{0.023} & 0.025      \\ \hline
    D4  &  \textbf{0.073}   &  0.270 & 0.475  &    0.215  & 0.211 & \underline{0.151}      \\ \hline
    D5  &  \textbf{0.152}   &  0.315 & 0.577  &    0.386  & 0.267 & \underline{0.165}     \\ \hline
    D6  &  \textbf{0.081}   &  0.310 & 0.540  &    0.458  & 0.193 & \underline{0.140}     \\ \hline
    D7  &  \textbf{0.092}   &  0.392 & 0.578  &    0.354  & 0.318 & \underline{0.189}     \\ \hline
    D8  &  0.061 &  \textbf{0.024} & 0.348  &    0.139  & \underline{0.068} & 0.069   \\ \hline
    D9  &  \underline{0.436} &  0.481 & 0.895  &    0.918  & 0.574 & \textbf{0.418}    \\ \hline
    D10  &  \textbf{0.087}  & 0.261  & 0.451  &    0.257  & 0.163 & \underline{0.096}   \\ \hline
    D11 &  \textbf{0.245}   &  0.723 & 0.859  &    0.823  & 0.621 & \underline{0.349}    \\ \hline
    D12  &  \textbf{0.023}  & 0.048 & 0.131  &    0.029  & 0.032 & \underline{0.026}  \\ \hline
    D13  &  \underline{0.068}   & 0.238 & 0.531  &    0.177  & 0.165 & \textbf{0.067}   \\ \hline
    D14  &  \textbf{0.395}   & 0.744 & 0.911  &    0.924     & 0.637  & \underline{0.414}     \\ \hline
    D15  &  \textbf{0.023}   & 0.053 & 0.083  &    0.045  & \underline{0.024}  & \underline{0.024}    \\ \hline
    D16  &  \textbf{0.270}   & 0.855 & 0.929  &  0.723   & 0.712  & \underline{0.456}   \\ \midrule[1pt]
    Average  &\textbf{0.158} &  0.364 & 0.551 &  0.438 &  0.301 & \underline{0.199}    \\ \bottomrule[1pt]
\end{tabular}
\end{table}
To validate the multi-dataset unified learning capability, we first evaluate the time-domain and frequency-domain prediction performance of WiFo across the 16 pre-training datasets.
WiFo is pre-trained on all training and validation samples of Dataset D1 to D16, with a total of 160k diverse CSI samples. 
Then, the pre-trained WiFo is evaluated on test samples of these 16 datasets.
In contrast, for deep learning-based approaches, the network is individually trained and tested on each dataset for either time-domain or frequency-domain prediction tasks.
The NMSE performance on the time-domain and frequency-domain task is illustrated in Table \ref{D1-D16-T} and Table \ref{D1-D16-F}, respectively.
Due to space limitations, only the performance of the base version of WiFo on each dataset is presented here, and the impact of the model size on performance is analyzed in Section \ref{scaling}.
For both the time-domain and frequency-domain prediction tasks, WiFo achieves the SOTA average NMSE performance and shows the best or the second-best prediction performance on most datasets.
The results show that WiFo can effectively perform unified learning across multiple datasets with different CSI configurations, while simultaneously mastering both time-domain and frequency-domain prediction capabilities.

\subsubsection{Zero-shot Generalization}
\begin{figure}[!t]
\centering
\subfloat[Time-domain channel prediction. ]{
		\includegraphics[scale=0.5]{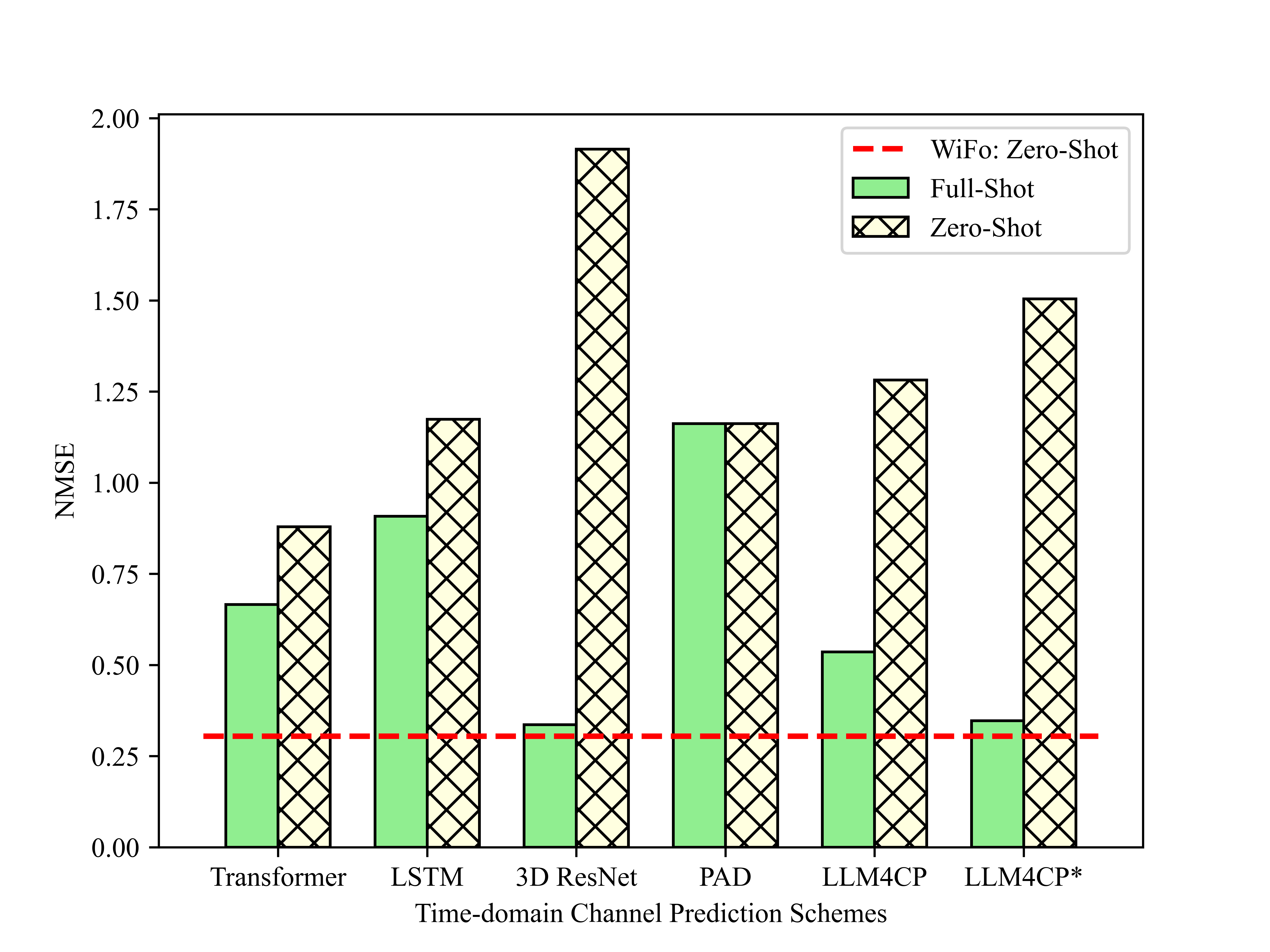}}
\subfloat[Frequency-domain channel prediction. ]{
		\includegraphics[scale=0.5]{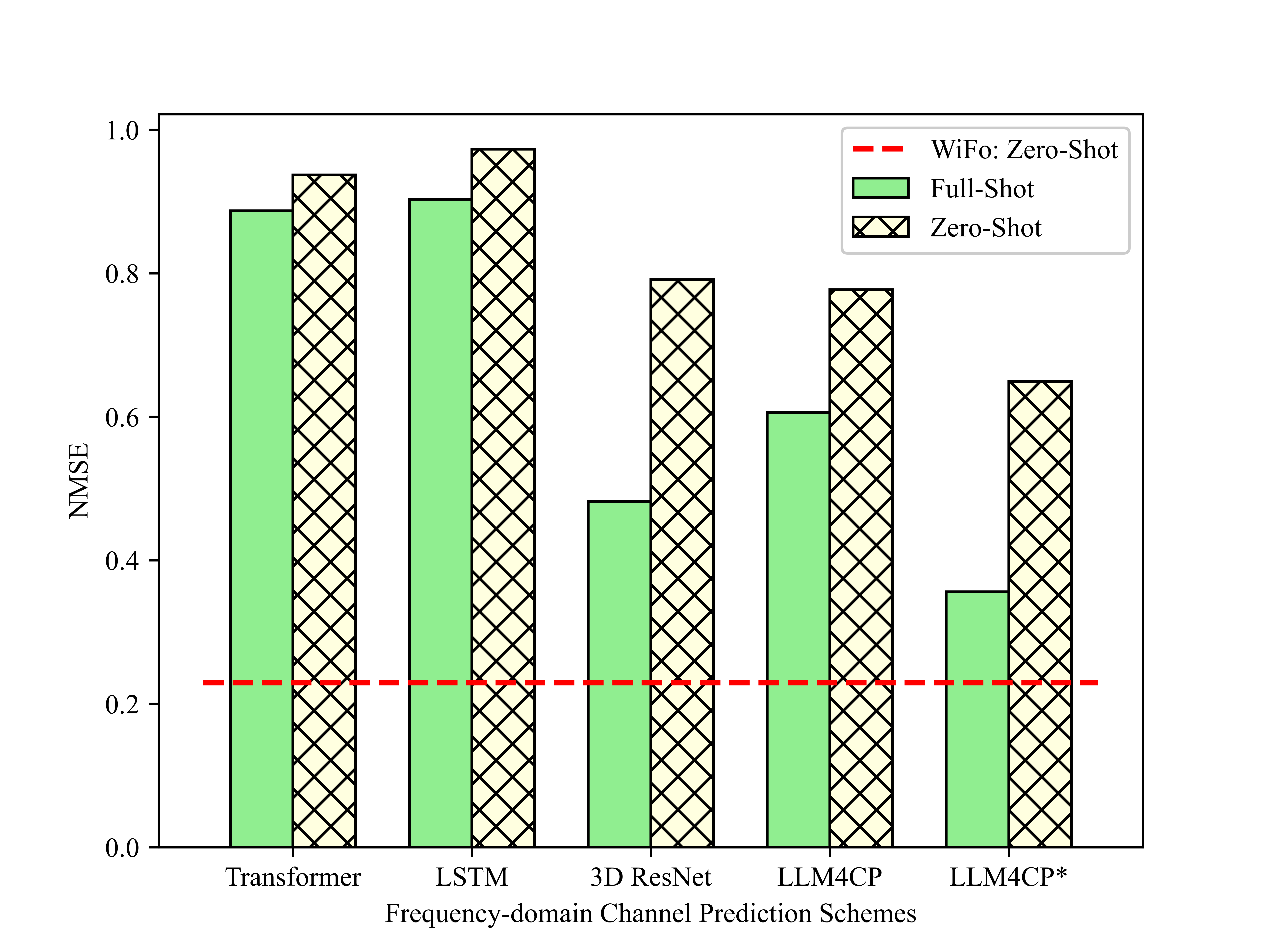}}
\caption{The zero-shot performance of WiFo and the full shot/zero-shot performance of other baselines on the D17 dataset.}\label{D17}
\end{figure}
\begin{figure}[!t]
\centering
\subfloat[Time-domain channel prediction. ]{
		\includegraphics[scale=0.5]{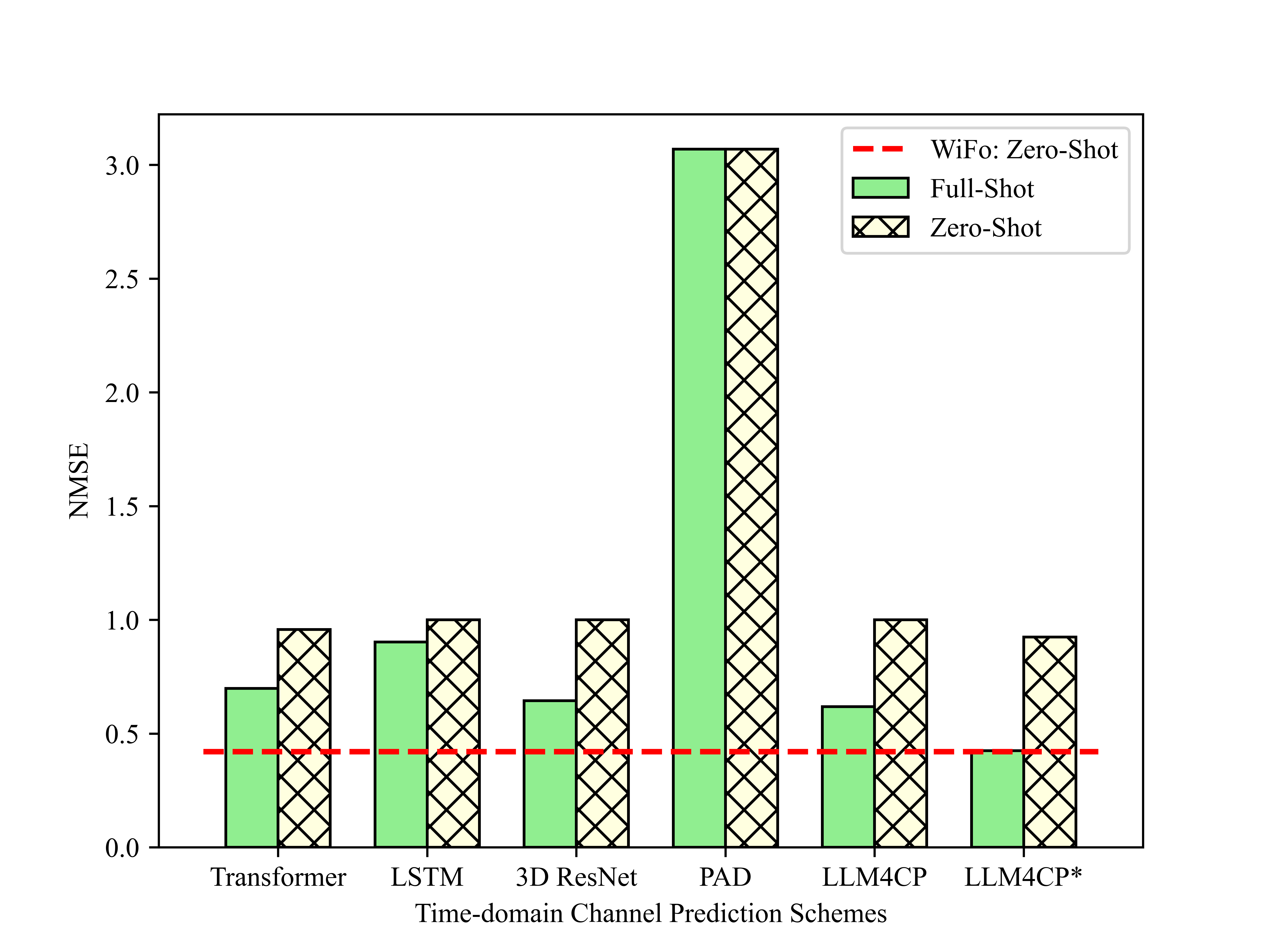}}
\subfloat[Frequency-domain channel prediction. ]{
		\includegraphics[scale=0.5]{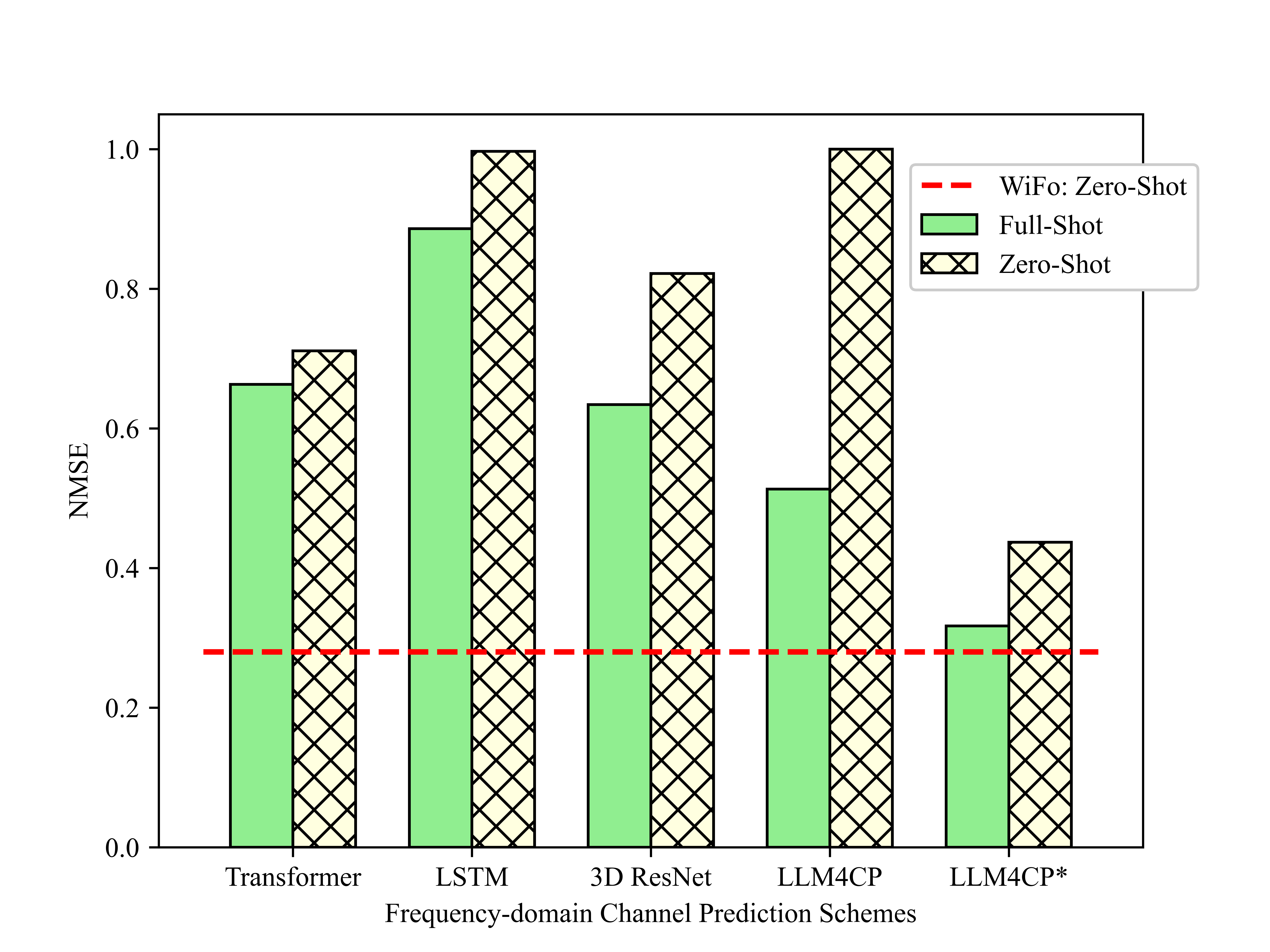}}
\caption{The zero-shot performance of WiFo and the full shot/zero-shot performance of other baselines on the D18 dataset.}\label{D18}
\end{figure}
\begin{figure}[!t]
\centering
\subfloat[Time-domain channel prediction. ]{
		\includegraphics[scale=0.5]{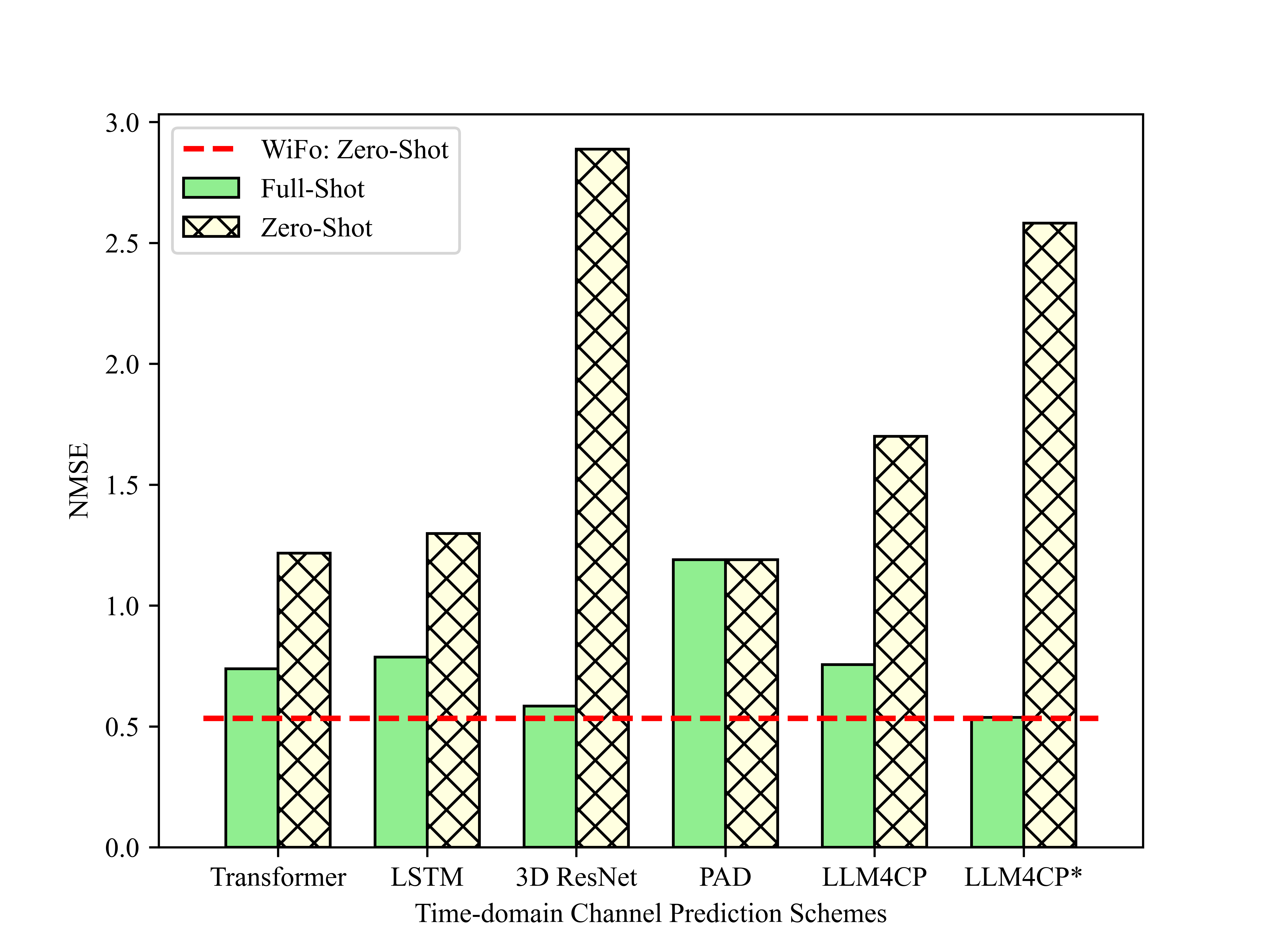}}
\subfloat[Frequency-domain channel prediction. ]{
		\includegraphics[scale=0.5]{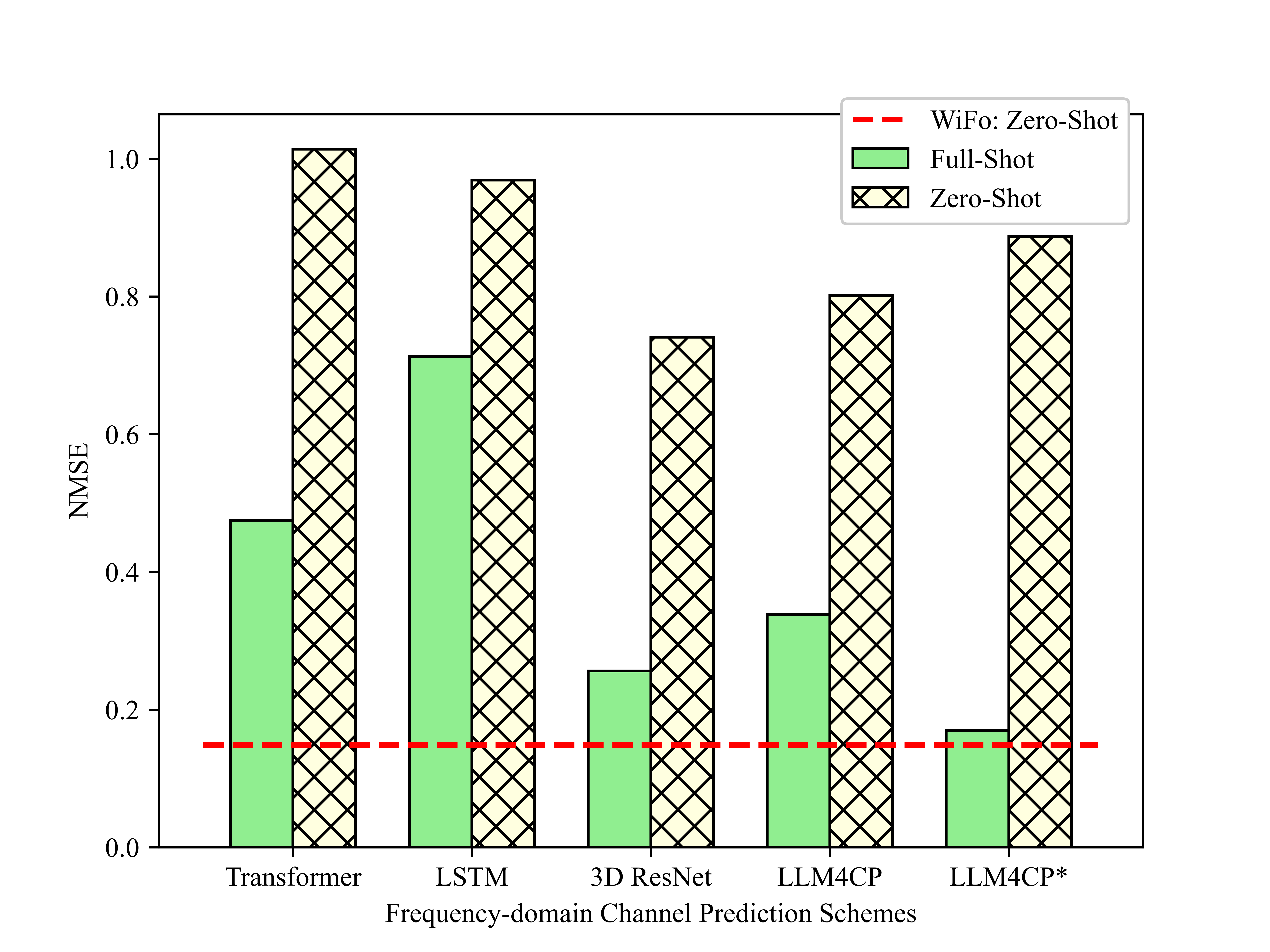}}
\caption{The zero-shot performance of WiFo and the full shot/zero-shot performance of other baselines on the D19 dataset.}\label{D19}
\end{figure}
To evaluate the generalization ability of WiFo, its zero-shot prediction capability is evaluated.
Specifically, the pre-trained WiFo performs inference directly on unseen datasets without any fine-tuning.
The D17, D18, and D19 datasets are used for zero-shot inference because their operating carrier frequencies are not included in the training set.
For other deep learning-based baselines, we consider both zero-shot and full-shot learning scenarios.
{For zero-shot learning, given that these methods struggle to generalize across CSI with varying shapes, they are trained on the D7, D13, and D7 datasets and then perform inference on the D17, D18, and D19 datasets, respectively.}
For full-shot learning, these methods are trained and tested on the same dataset. 
{The performance of WiFo-Base and other baselines on the D17, D18, and D19 datasets is illustrated in Fig. \ref{D17}, Fig. \ref{D18}, and Fig. \ref{D19}, respectively.}

It is observed that most deep learning-based methods struggle to perform zero-shot inference.
Additionally, the transformer and LSTM show poor full-shot performance and even fail to learn, indicating that prediction on D17, D18, and D19 datasets is highly challenging.
This is because the three datasets have large sub-carrier intervals and a large number of antennas, making it difficult for these models to learn the 3D variations of the CSI.
Nevertheless, the zero-shot performance of WiFo outperforms the zero-shot and even full-shot performance of all other methods, demonstrating its remarkable cross-frequency generalization ability.
{Its superiority in zero-shot prediction can be attributed to the pre-training paradigm, where WiFo learns a generalizable channel prediction capability applicable to various CSI distributions through training on vast heterogeneous datasets.}
Therefore, once WiFo is trained on large-scale datasets, it can potentially be deployed instantly at the BS, significantly reducing the costs associated with data collection and model fine-tuning.

\subsubsection{Ablation Study}

\begin{table}[t]
\footnotesize
\caption{Results of ablation experiments. The best results are highlighted in \textbf{bold}, while the second-best results are \underline{underlined}.}\label{ablation}
\centering
\begin{tabular}{cccccccc}
\toprule[1pt]
      & \multicolumn{3}{c}{\textbf{Time-domain prediction}} & \multicolumn{3}{c}{\textbf{Frequency-domain prediction}} & \\\cline{2-7}
                                        & D1-D16         & D17         & D18         & D1-D16         & D17             & D18   & \textbf{Average}      \\ \midrule[1pt]
WiFo-Base                               &  \underline{0.210}  & \textbf{0.305}     & \underline{0.420}      & {0.158}  & {0.229}        & {0.280}     & \textbf{0.267}       \\ \hline
w/ learnable PE\cite{feichtenhofer2022masked}                         &  0.224        & 0.354      &  0.442    &   \underline{0.154}          &  \underline{0.220}       &   \underline{0.267}    & 0.277  \\ \hline
w/o random-masked reconstruction        &  0.214        & 0.317       &  0.435    &  0.165           & 0.233              & 0.294   & \underline{0.276}  \\ \hline
w/o time-masked reconstruction          &  0.497        & 0.754       &  0.723    &  \textbf{0.145}           & \textbf{0.199}              & \textbf{0.257}  & 0.429   \\ \hline
w/o frequency-masked reconstruction    &  \textbf{0.205}        & \underline{0.310}       &  \textbf{0.412}     &  0.472           & 0.819              & 0.656  & 0.479   \\ \bottomrule[1pt]
\end{tabular}
\end{table}
To verify the effectiveness of the specialized designs in WiFo, we perform ablation studies based on the WiFo-Base model.
The ablation results are shown in Table \ref{ablation}, where the corresponding multi-dataset unified learning and zero-shot generalization performance are given.
The average performance is measured by the mean NMSE of columns 2 to 7, representing overall performance.
Replacing the proposed STF-PE with learnable space-time positional encoding designed for video would degrade overall performance.
{This can be attributed to the fact that the proposed STF-PE is an absolute positional encoding \cite{yuan2024unist}, which is better suited to token sequences of varying lengths compared to learnable positional encodings.}
Additionally, removing the random-masked reconstruction task degrades the unified learning and generalization performance for both the time-domain and frequency-domain channel prediction, highlighting its effectiveness for self-supervised pre-training.
It can be attributed that the additional random masking strategy facilitates WiFo's ability to capture the intrinsic 3D relationships of CSI.
Moreover, removing time-masked or frequency-masked reconstruction enhances frequency-domain or time-domain channel prediction performance but significantly degrades the other.
Therefore, both time-masked and frequency-masked reconstruction tasks are essential for the model pre-training to address both prediction tasks simultaneously.

\subsubsection{Scaling Analysis}\label{scaling}

\begin{table}[t]
\centering\footnotesize
\caption{The performance of WiFo is evaluated across different model sizes and various pre-training dataset scales. The best results are highlighted in \textbf{bold}, while the second-best results are \underline{underlined}.}\label{scaling table}
\begin{tabular}{cccccccc}
\toprule[1pt]
                        &             & \multicolumn{3}{l}{Time-domain Prediction} & \multicolumn{3}{l}{Frequency-domain Prediction} \\ \cline{3-8} 
Pre-training Dataset    & Model       & D1,D5,D9,D15         & D17         & D18         & D1,D5,D9,D15            & D17           & D18          \\ \midrule[1pt]
\multirow{5}{*}{D1-D16} & WiFo-Tiny   &  0.315           &  0.444       &  0.506     & 0.343       &  0.440       &  0.399            \\ \cline{2-8} 
                        & WiFo-Little &  0.271        &  0.371       & 0.446      & 0.260       &  0.284       &  0.301            \\ \cline{2-8} 
                        & WiFo-Small  &  0.245        & 0.326        &  0.421     & 0.239       & 0.245        &  0.299            \\ \cline{2-8} 
                        & WiFo-Base   &  \underline{0.237}        & \textbf{0.305}        & \underline{0.420}      & \underline{0.232}       & \textbf{0.229}        & \underline{0.280}       \\ \cline{2-8} 
                        & WiFo-Large  & \textbf{0.234}         &  \underline{0.314}       & \textbf{0.416}      & \textbf{0.226}       &  \underline{0.232}       &  \textbf{0.276}            \\ \hline
D1,D2,D5,D6,D9,D12,D15,D16& WiFo-Base & 0.261         & 0.356        &  0.477     & 0.258       & 0.269        &  0.345            \\ \hline
D1,D5,D9,D15              & WiFo-Base & 0.306         & 0.960        & 0.636      & 0.320       & 1.036        & 0.523             \\ \bottomrule[1pt]
\end{tabular}
\end{table}
Scaling analysis is essential for foundation models as it highlights the effects of model size and pre-training dataset scale on performance. In our scaling experiments, we consider five model sizes and three dataset scales to evaluate their overall performance. We assess unified learning performance using the average prediction NMSE across the D1, D5, D9, and D15 datasets, and evaluate zero-shot performance on the D17 and D18 datasets, as presented in Table \ref{scaling table}.

From the first five rows of Table \ref{scaling table}, we observe that, with a fixed pre-training dataset scale, increasing the model size generally enhances both the unified learning performance and zero-shot generalization capabilities within the observed range. This improvement is attributed to larger models' ability to capture more complex patterns, resulting in better learning. However, once the model reaches a certain size, its zero-shot generalization ability plateaus, constrained by the scale of the pre-training dataset.
{It is due to the fact that as the model parameters increase, WiFo gradually overfits the pre-training dataset, leading to performance degradation on out-of-distribution datasets.}

Conversely, as seen in rows 4, 6, and 7, expanding the pre-training dataset scale significantly boosts both unified learning performance and zero-shot performance simultaneously. 
In summary, the unified learning performance and generalization performance of WiFo are improved \cite{kaplan2020scaling} with both the increased model size and the scale of the pre-training datasets.
Therefore, WiFo is a scalable model that has the potential for further improvement as computational power and dataset scale increase. 
\subsubsection{Network Storage and Inference Cost}
\begin{table}[t]
\footnotesize
\caption{{The number of parameters, FLOPs, and inference time per batch (batch size is set to 8) of each model.}}\label{cost}
\centering
\begin{tabular}{cccccccccc}
\toprule[1pt]
                   & WiFo-Base & Transformer & LSTM       & 3D ResNet & PAD   & LLM4CP     & LLM4CP*  \\ \midrule[1pt]
Parameters(M)      & 21.60      & 0.91        & 1.13       & 31.73    &  0     & 82.35      & 83.32     \\ \hline
{FLOPs(G)}            & {1.456}      & {0.211}       & {0.061}      & {390.560}   & {0.003}      & {1.720}      & {11.520}     \\ \hline
Inference time(ms) & 9.659      & 6.238       & 5.209      & 86.865   & 119.397& 4.718      & 7.342    \\ \bottomrule[1pt]
\end{tabular}
\end{table}
{The number of parameters, floating point operations (FLOPs), and inference time of models are closely tied to their storage and computational overhead, directly impacting the practical deployment of channel prediction models at the BS.}
These overhead metrics are shown in Table \ref{cost}, where the batch size is set as 8 for inference time calculation, and inference samples are taken from dataset D1.
{It is worth noting that inference time is not solely determined by FLOPs but also depends on the network architecture and hardware optimization.
As a model-based method, PAD has a lower storage cost and FLOPs but a longer inference time due to the lack of GPU optimization.}
For 3D ResNet, complex 3D convolution operations and the large number of parameters result in significant inference overhead.
For LLM4CP, its antenna-parallelized version has a higher inference time because the number of CSI samples processed per batch increases proportionally with the number of antennas.
It is observed that WiFo has an acceptable parameter count and comparable inference time among these channel prediction schemes.

{Moreover, it is worth noting that WiFo is a versatile model capable of replacing multiple specialized channel prediction models, which can significantly reduce the number of models required on the BS.
For instance, in the experiment introduced in Section 4.3.1, a single WiFo model is sufficient to perform time-domain and frequency-domain prediction tasks across 16 datasets, whereas other deep learning-based approaches require training 32 separate models for the above tasks.
The parameter count of WiFo-Base is even lower than the total parameters of 32 transformer models, thereby reducing additional deployment overhead.}

\section{Conclusions}
In this paper, we have introduced a novel wireless foundation model, WiFo, designed to simultaneously facilitate time-domain and frequency-domain channel prediction tasks as well as diverse 3D CSI configurations. We developed an MAE-based network structure and implemented several mask reconstruction tasks for self-supervised pre-training to capture the intrinsic space-time-frequency features of CSI. WiFo, pre-trained on large-scale heterogeneous datasets, can be directly deployed for inference without any need for fine-tuning. Simulations demonstrate its exceptional performance in unified multi-dataset training and its superb zero-shot generalization capabilities with acceptable  inference overhead.






\begin{thebibliography}{99}

\bibitem{cheng2023intelligent}Cheng X, Zhang H, Zhang J, et al. {Intelligent multi-modal sensing-communication integration: synesthesia of machines}. IEEE Commun Surveys Tuts, 2024, 26: 258-301
\bibitem{gao2020estimating} Gao S, Cheng X, Yang L. Estimating doubly-selective channels for hybrid mmWave massive MIMO systems: A doubly-sparse approach. IEEE Trans Wirel Commun, 2020, 19: 5703-5715.
\bibitem{yin2020addressing}Yin H, Wang H, Liu Y, et al. Addressing the curse of mobility in massive MIMO with prony-based angular-delay domain channel predictions. IEEE J Select Areas Commun, 2020, 38: 2903-2917
\bibitem{rottenberg2020performance}Rottenberg F, Choi T, Luo P, et al. Performance analysis of channel extrapolation in FDD massive MIMO systems. IEEE Trans Wirel Commun, 2020, 19: 2728-2741
\bibitem{wang2022transformer}Wang Y, Gao Z, Zheng D, et al. Transformer-empowered 6G intelligent networks: From massive MIMO processing to semantic communication. IEEE Wirel Commun, 2022, 30: 127-135.
\bibitem{zhu2023pushing}Zhu G, Lyu Z, Jiao X, et al. Pushing AI to wireless network edge: An overview on integrated sensing, communication, and computation towards 6G. Sci China Inf Sci, 2023, 66: 130301.
\bibitem{jiang2022accurate}Jiang H, Cui M, Ng D W K, et al. Accurate channel prediction based on transformer: making mobility negligible. IEEE J Select Areas Commun, 2022, 40: 2717-2732
\bibitem{liu2022spatio}Liu G, Hu Z, Wang L, et al. Spatio-temporal neural network for channel prediction in massive MIMO-OFDM systems. IEEE Trans Commun, 2022, 70: 8003-8016
\bibitem{liu2021fire}Liu Z, Singh G, Xu C, et al. FIRE: enabling reciprocity for FDD MIMO systems. In: Proceedings of the 27th Annual International Conference on Mobile Computing and Networking, New York, 2021. 628-641
\bibitem{zhang2024integrated} Zhang H, Gao S, Cheng X, et al. Integrated sensing and communications towards proactive beamforming in mmWave V2I via multi-modal feature fusion (MMFF). IEEE Trans Wirel Commun, 2024, 23: 15721-15735
\bibitem{liu2024llm4cp}Liu B, Liu X, Gao S, et al. LLM4CP: Adapting Large Language Models for Channel Prediction. J Commun Inf Netw, 2024, 9: 113-125
\bibitem{bommasani2021opportunities}Bommasani R, Hudson D A, Adeli E, et al. On the opportunities and risks of foundation models. 2021. ArXiv:2108.07258
\bibitem{fontaine2024towards}Fontaine J, Shahid A, De Poorter E. Towards a Wireless Physical-Layer Foundation Model: Challenges and Strategies. 2024. ArXiv: 2403.12065
\bibitem{salihu2024self}Salihu A, Rupp M, Schwarz S. Self-Supervised and Invariant Representations for Wireless Localization. IEEE Trans Wirel Commun, 2024, 23: 8281-8296
\bibitem{alikhani2024large}Alikhani S, Charan G, Alkhateeb A. Large Wireless Model (LWM): A Foundation Model for Wireless Channels. 2024. ArXiv: 2411.08872
\bibitem{zhang2024latest}Zhang Y, Zhang J, Pei Y, et al. Latest progress for 3GPP ISAC channel modeling standardization. Sci China Inf Sci, 2024, 67: 217301
\bibitem{he2022masked}He K, Chen X, Xie S, et al. Masked autoencoders are scalable vision learners. In: Proceedings of the IEEE/CVF conference on computer vision and pattern recognition, New Orleans, 2022. 16000-16009
\bibitem{tong2022videomae}Tong Z, Song Y, Wang J, et al. Videomae: Masked autoencoders are data-efficient learners for self-supervised video pre-training. In: Advances in neural information processing systems, New Orleans,2022. 10078-10093
\bibitem{feichtenhofer2022masked}Feichtenhofer C, Li Y, He K. Masked autoencoders as spatiotemporal learners. Advances in neural information processing systems, 2022, 35: 35946-35958.
\bibitem{dosovitskiy2020image}Alexey D. An image is worth 16x16 words: Transformers for image recognition at scale. 2020. ArXiv: 2010.11929
\bibitem{jaeckel2014quadriga}Jaeckel S, Raschkowski L, Börner K, et al. QuaDRiGa: A 3-D multi-cell channel model with time evolution for enabling virtual field trials. IEEE Trans Antennas Propagat, 2014, 62: 3242-3256.
\bibitem{3gpp2018study}3GPP Radio Access Network Working Group. Study on channel model for frequencies from 0.5 to 100 GHz (Release 15). 3GPP TR 38.901, 2018.
\bibitem{feichtenhofer2019slowfast}Feichtenhofer C, Fan H, Malik J, et al. Slowfast networks for video recognition. In: Proceedings of the IEEE/CVF international conference on computer vision, Seoul, 2019. 6202-6211.
\bibitem{jiang2020deep}Jiang W, Schotten H D. Deep learning for fading channel prediction. IEEE Open J Commun Soc, 2020, 1: 320-332.
\bibitem{loshchilov2016sgdr}Loshchilov I, Hutter F. Sgdr: Stochastic gradient descent with warm restarts. 2016. ArXiv: 1608.03983
\bibitem{yuan2024unist}Yuan Y, Ding J, Feng J, et al. Unist: A prompt-empowered universal model for urban spatio-temporal prediction. In: Proceedings of the 30th ACM SIGKDD Conference on Knowledge Discovery and Data Mining, New York, 2024. 4095-4106.
\bibitem{kaplan2020scaling}Kaplan J, McCandlish S, Henighan T, et al. Scaling laws for neural language models. 2020. ArXiv: 2001.08361

\end{thebibliography}
\end{document}